\documentclass[twocolumn,aps,superscriptaddress,prl]{revtex4}
\usepackage{graphicx}
\usepackage{times}
\usepackage{amsmath}
\usepackage{amssymb}
\usepackage{units}
\usepackage{stmaryrd} 

\renewcommand{\vec}[1]{\boldsymbol{#1}}

\begin{document}
\preprint{0}

\title{Ag-coverage-dependent symmetry of the electronic states of the Pt(111)-Ag-Bi interface:\\ the ARPES view of a structural transition }

\author{E. Frantzeskakis}
\thanks{Current address: Synchrotron SOLEIL, L'Orme des Merisiers, Saint Aubin-BP 48, 91192 Gif sur Yvette Cedex, France}
\email{frantzeskakis@synchrotron-soleil.fr}
\affiliation{Institute of Condensed Matter Physics (ICMP), Ecole Polytechnique
F{\'e}d{\'e}rale de Lausanne (EPFL), Station 3, CH-1015 Lausanne,
Switzerland}

\author{S. Pons}
\affiliation{Institute of Condensed Matter Physics (ICMP), Ecole Polytechnique
F{\'e}d{\'e}rale de Lausanne (EPFL), Station 3, CH-1015 Lausanne,
Switzerland}\affiliation{Institut des NanoSciences de Paris (INSP) , Universit{\'e} Pierre et Marie Curie (UPMC) - Paris 6 - CNRS, France}

\author{A. Crepaldi}
\affiliation{Institute of Condensed Matter Physics (ICMP), Ecole Polytechnique
F{\'e}d{\'e}rale de Lausanne (EPFL), Station 3, CH-1015 Lausanne,
Switzerland}

\author{H. Brune}
\affiliation{Institute of Condensed Matter Physics (ICMP), Ecole Polytechnique
F{\'e}d{\'e}rale de Lausanne (EPFL), Station 3, CH-1015 Lausanne,
Switzerland}

\author{K. Kern}
\affiliation{Institute of Condensed Matter Physics (ICMP), Ecole Polytechnique
F{\'e}d{\'e}rale de Lausanne (EPFL), Station 3, CH-1015 Lausanne,
Switzerland}\affiliation{Max-Planck-Institut f\"{u}r
Festk\"{o}rperforschung, D-70569 Stuttgart, Germany}

\author{M. Grioni}
\affiliation{Institute of Condensed Matter Physics (ICMP), Ecole Polytechnique
F{\'e}d{\'e}rale de Lausanne (EPFL), Station 3, CH-1015 Lausanne,
Switzerland}

\date{\today}

\begin{abstract}
We studied by angle-resolved photoelectron spectroscopy the strain-related structural transition from a pseudomorphic monolayer (ML) to a
striped incommensurate phase in an Ag thin film grown on Pt(111).
We exploited the surfactant properties of Bi to grow ordered Pt(111)$-x$MLAg$-$Bi trilayers with $0 \le x \le 5$ ML, and monitored the dispersion of 
the Bi-derived interface states to probe the structure of the underlying Ag film. We find that their symmetry changes from threefold to sixfold and 
back to threefold in the Ag coverage range studied. Together with previous Scanning Tunneling Microscopy and photoelectron diffraction data, these 
results provide a consistent microscopic description of the coverage-dependent structural transition. 
\end{abstract}

\maketitle

\section{I. Introduction}

Lattice mismatch is a crucial factor determining the growth mode and morphology of heteroepitaxial metal-metal interfaces.
It gives rise to pseudomorphic strained layers, but it can also be accommodated by the formation of either moir\'e structures
or incommensurate phases containing misfit dislocations where the
strain is locally relieved. Strain can also be important in homoepitaxial systems. The Au(111) surface is a paradigm of
the latter. A $4$\% strain in the topmost layer is relieved by
the formation of a pairwise dislocation network, yielding the well-studied ($\sqrt{3}\times 22$)
herringbone reconstruction \cite{Tanishiro1981,Barth1990}.

The Pt(111)-Ag interface is a typical example of strain relief in a heteroepitaxial
system \cite{RoderPRL1993,Roder1993,Brune1994,Roder1997}. For sub-monolayer (ML) coverages it exhibits partial dislocations, which are removed 
by annealing to $800$ K, or at the completion of the first ML at RT (a ``re-entrant pseudomorphic growth''). The first complete Ag ML is
compressed with respect to a bulk Ag(111) plane. The strain is
relieved in the second Ag ML by the formation of a metastable striped
incommensurate (SI) phase at room temperature. The SI phase transforms to an equilibrium structure with a triangular
dislocation network above $800$ K. In both phases dislocation lines
separate domains with fcc and hcp stacking
\cite{Rangelov1995,Bromann1997}. 

The structure and properties of this interface
depend on the Ag coverage. Therefore, in order to perform consistent studies, one needs to calibrate the
amount of Ag on the surface. While the morphology and structure of Pt-Ag(111) have been thoroughly investigated, relatively little is known of its
microscopic electronic properties. We present here angle-resolved photoelectron spectroscopy (ARPES) data on the 
band structure of Ag-Pt(111) and of a Pt(111)$-x$MLAg$-$Bi trilayer system 
with $1 \le x \le 5$ ML. The latter was suggested by recent experiments on a BiAg$_2$ surface alloy grown on Ag(111), showing a 
very large separation of opposite spin states (Rashba-Bychkov effect) \cite{Ast2007,Meier2008}. Theory predicts \cite{Bihlmayer2007} that 
the size of the Rashba-Bychkov (RB) effect  is very sensitive to slight changes in the atomic structure, motivating us to explore the 
possible influence of interfacial strain on the spin-orbit splitting.

Here we show that rather than the  $(\sqrt{3}\times\sqrt{3})$R30$^{\circ}$ BiAg$_2$ surface alloy formed on the Ag(111) substrate, Bi atoms arrange 
themselves in an ordered overlayer with a $(2\times2)$  symmetry. The resulting  band structure is distinct from that of the alloy and does not exhibit signatures
of a large spin-orbit splitting. By contrast, the Bi-derived states effectively probe the structure of the Ag film. Their angular dispersion is 
determined by the symmetry of the underlying layer. It exhibits a change from threefold to sixfold in correspondence of the structural transition to the 
SI phase at $2$ ML coverage, and then back to threefold for larger Ag thicknesses. These observations support the general model of the transition 
proposed on the basis of Scanning Tunneling Microscopy (STM) and photoelectron diffraction studies \cite{Brune1994,Rangelov1995,Roder1997}.

\section{II. Experimental Details}

The Pt(111) substrate was prepared by repeated cycles of Ar sputtering and
annealing at $1300$ K. The crystal was then exposed to an O$_{2}$
partial pressure of $P\ =1\times10^{-7}$mbar at $900$ K, in order to catalytically remove
the carbon impurities which had segregated from the bulk. Finally, it was annealed at $1000$ K without O$_{2}$.
The order and cleanliness of the surface were verified by means of LEED and ARPES.

Ag was evaporated from a resistively heated tungsten basket which had been
accurately calibrated in previous experiments
\cite{Frantzeskakis2008,FrantzeskakisCrep2010}. Bi was deposited
by electron-beam-assisted evaporation using a commercial EFM3
Omicron source. The sample was kept at room temperature (RT)
during the deposition of both Ag and Bi. A mild post-annealing resulted in sharp
LEED spots. The deposition order of Ag and
Bi could be reversed without any effect on the crystalline order and the symmetry and the
electronic states, as probed respectively by LEED and ARPES.

ARPES spectra were acquired at RT and $21.2$ eV photon energy using
a Phoibos $150$ Specs Analyzer equipped with a monochromatized
Gammadata VUV $5000$ high brightness source. The ultimate resolutions of
the experimental setup are $5$ meV (energy) and $0.2^{\textmd{o}}$ (angular). In the present
work, broad $k$-range ARPES spectra were acquired by a sequential scanning of the
polar angle. The angular step was $0.5^{\textmd{o}}$. 
This is superior than the FWHM of the sharpest state measured in the studied interface ($0.11$ \ \AA$^{-1}$).
The corresponding experimental energy resolution was set to around $40$ meV. During the stepwise scanning of the
polar angle, the incidence angle varies from $45^{\textmd{o}}$ (at normal emission)
towards normal incidence at higher $k$ values. The HeI source is partially polarized ($80$\% $\sigma$-polarized)
due to two reflections. The base pressure was in the low $10^{-10}$\ mbar range and increased up to $10^{-9}$\ mbar
during measurements due to He gas leakage from the discharge cavity.

\begin{figure}[!b]
  \centering
  \includegraphics[width = 7.9 cm]{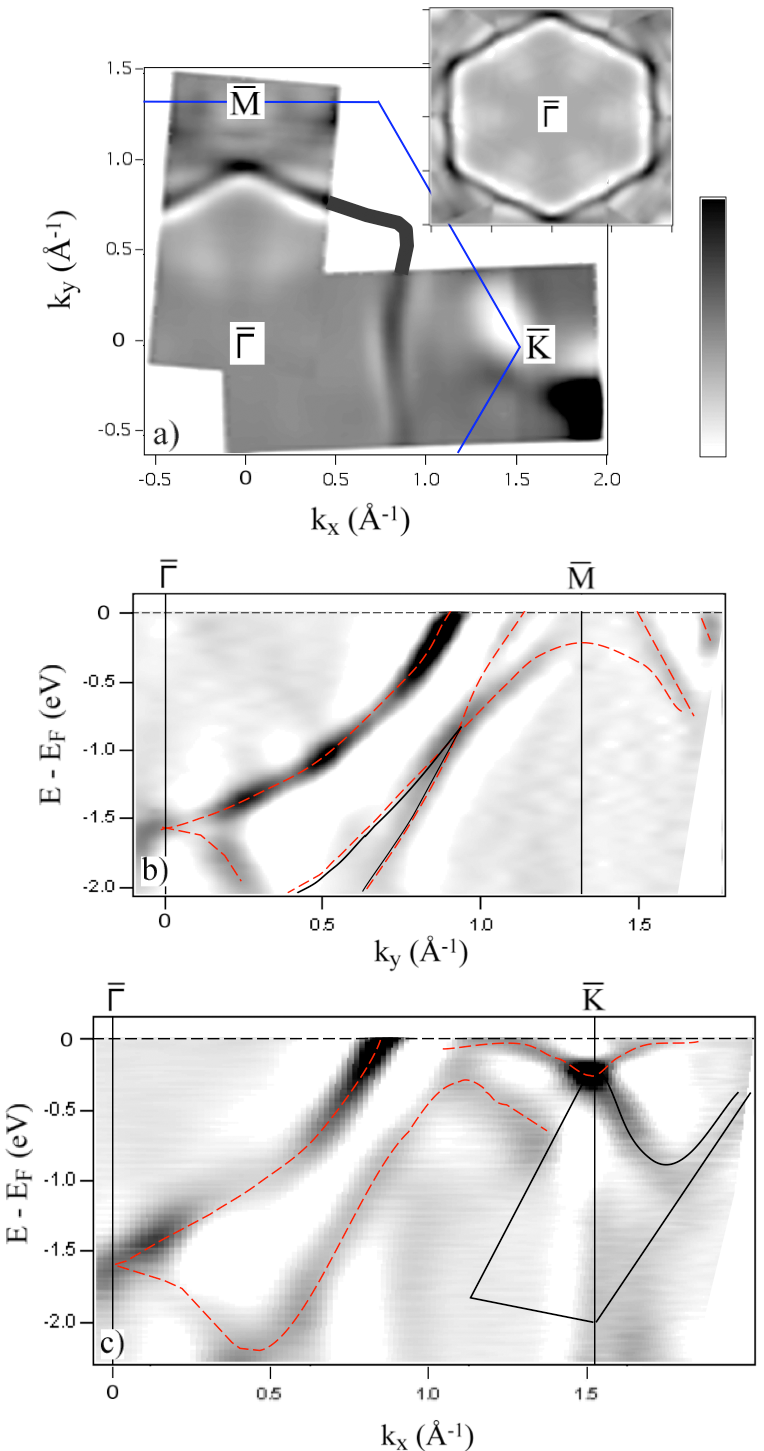}
  \caption{(color online) ARPES results for the Pt(111) substrate. (a) Constant energy ARPES intensity map at E=E$_{\textmd{F}}$. Thin solid lines follow the
 Pt(111) $(1\times1)$ surface Brillouin zone. The thick solid line is a guide to the eye depicting the constant energy contour in a $k$ space region where
 no measurement was performed. The inset presents the sixfold symmetric surface resonance (see text).
(b) and (c) ARPES intensity plots illustrate the band dispersion along the $\overline{\Gamma \textmd{M}}$ and $\overline{\Gamma
\textmd{K}}$ high-symmetry directions. Dashed curves are guides to the eye and solid lines highlight the projected bulk band gaps
\cite{Di1992,Ramstad2000}. The second derivative of the photoemission intensity has been used to enhance the experimental features.
Intensity follows the attached gray-scale bar where signal-to-noise ratio increases from the bottom to the top.} \label{fig1}
\end{figure}

\begin{figure}[!b]
  \centering
  \includegraphics[width = 8.5 cm]{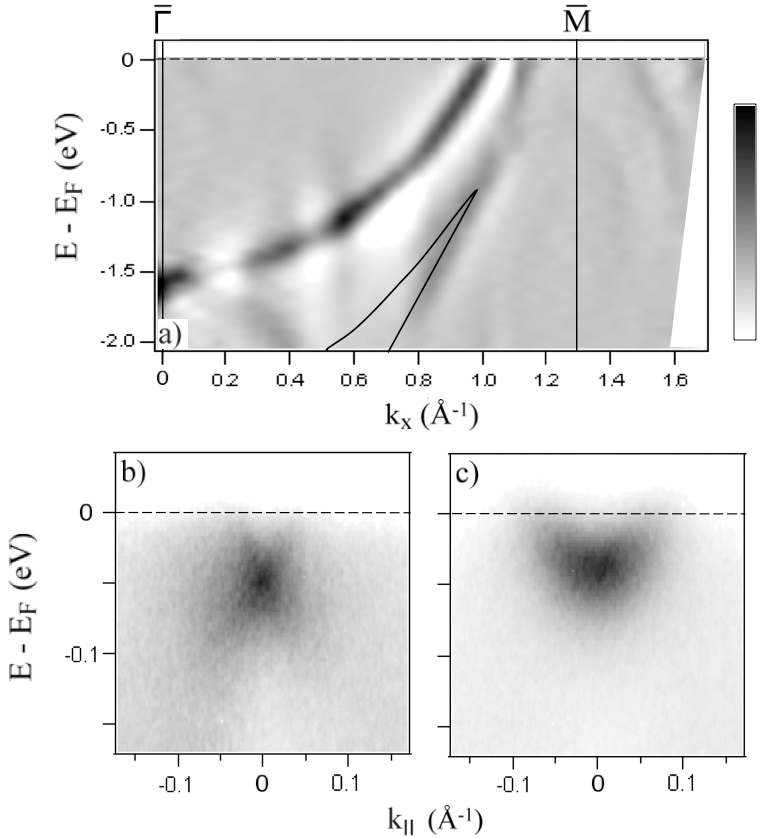}
  \caption{Pt(111)$-x$MLAg: (a) Electronic band dispersion along $\overline{\Gamma \textmd{M}}$ for $x=1$.
There is no major difference from the band structure of clean Pt(111) [i.e. Fig. 1(b)]. (b),(c) Band dispersion around $\overline{\Gamma}$
when $x$ is around $3$ ML (b) and $5$ ML (c). The x-like feature in (b) is very different
than the parabolic Shockley Ag(111) surface state in (c) and its onset marks the $2$ ML coverage. The second 
derivative of the photoemission intensity has been used in (a) to enhance the experimental features.
In all cases, intensity follows the attached gray-scale bar where signal-to-noise ratio increases from the bottom to the top.} \label{fig2}
\end{figure}

\section{III. Results}

Figure 1 summarizes the ARPES results for the clean Pt(111)
surface. Data were collected in the region of $k$ space presented in Fig. 1(a). It shows a constant energy (CE) intensity map
measured at the Fermi energy (E$_{\textmd{F}}$). Figure 1(b) and 1(c) illustrate the experimental energy-wave vector
dispersion along the two high-symmetry directions $\overline{\Gamma \textmd{M}}$ and, respectively,
$\overline{\Gamma \textmd{K}}$ of the surface Brillouin zone (BZ). The BZ boundaries are at
$1.31$ \AA$^{-1}$ ($\overline{\Gamma \textmd{M}}$) and $1.51$ \AA$^{-1}$ ($\overline{\Gamma \textmd{K}}$).
Both the CE map and the band dispersion are dominated by a state centered  at the $\overline{\Gamma}$ point
and dispersing through E$_{\textmd{F}}$. It gives rise to a nearly hexagonal Fermi surface (FS).
The corner of the FS, along $\overline{\Gamma \textmd{M}}$, is 
at $k_{\textmd{F}}(\overline{\Gamma \textmd{M}})\simeq0.9$\ \AA$^{-1}$, while the Fermi crossing along 
$\overline{\Gamma \textmd{K}}$ is at $k_{\textmd{F}}(\overline{\Gamma \textmd{K}})\simeq0.8$\ \AA$^{-1}$

Hexagonal contours are expected for surface states at the (111) surfaces of fcc crystals, by contrast with 
the threefold symmetry of bulk states. This point is further discussed in Section IV. However the bulk electronic structure
of Pt does not present a projected gap around $\overline{\Gamma}$ that could support a surface state.
Indeed, when the photon energy is varied, this state exhibits a weak but finite $k_{\perp}$ dispersion, typical of a bulk state.
The hexagonal contour was then tentatively ascribed not to a true surface state, but to a surface resonance associated with the sixth bulk
band \cite{Di1992,Di1991}. That suggestion was later supported by a density functional theory (DFT) calculation
that found a state with $5d_{xz,yz}$ character and a strong ($10$\%) localization in the surface layer
\cite{Wiebe2005}. However, it should also be noted that the predicted ARPES FS contour, which has a threefold symmetry for a generic
photon energy, becomes nearly hexagonal for specific values of h$\nu$ (CE maps were calculated with the FLAN software, courtesy of 
Dr. E. Garcia-Michel \cite{Joco}). The transition from an almost sixfold to a clearly threefold
contour may occur by changing h$\nu$ - or, equivalently, the inner potential V$_0$ -- by less than $2$ eV,
so that the observation of a nearly hexagonal shape could be at least partly accidental. In this frame, the inset of Fig. 1(a), which
has been obtained by a sixfold symmetrization (same procedure followed in Ref. \onlinecite{Di1991}) should be solely considered
for a better visualization of the aforementioned surface contour and the sixfold structure closer to $\overline{\Gamma}$ which was previously 
associated with the fifth bulk band \cite{Di1991}. The sixfold symmetrization cannot be extended to higher $k$ values due to the
threeefold overall symmetry of the Pt(111) surface.

Two weaker band features are observed at larger wave vectors. 
Along $\overline{\Gamma\textmd{M}}$ one crosses E$_{\textmd{F}}$ at $\simeq1.1\ $\AA$^{-1}$ and
again at a symmetric point on the opposite side of $\overline{\textmd{M}}$, while
the second
has a maximum at $\overline{\textmd{M}}$. Their dispersion
follows the edges of the
projected bulk continuum, defined by solid lines in the figure, and
may continue as a surface resonance \cite{Di1992,Ramstad2000,MacDonald1981}. Along
the $\overline{\Gamma \textmd{K}}$ direction, there is a strong feature around the $\overline{\textmd{K}}$
point, where previous studies predicted the edge of a
bulk projected gap and a relatively flat surface resonance \cite{Di1992,Ramstad2000}.

The deposition of silver on the Pt(111) substrate was monitored by LEED and ARPES. For $\Theta_{\textmd{Ag}}=1$ ML
the LEED pattern and the band structure within the range of Fig. 1 [Fig. 2(a)] are essentially identical to those of 
the clean substrate. 
New features appear at the completion of the second ML, namely satellite spots pointing towards a  $(\sqrt{3}\times n)$ reconstruction,
characteristic of the SI phase \cite{Brune1994,Roder1997,Mansour}. 
Moreover, between $2$ ML and $4$ ML the ARPES intensity map [Fig. 2(b)] exhibits an ``x-like" feature just below E$_{\textmd{F}}$ centered at
the $\overline{\Gamma}$ point.  This
structure was recently observed in an independent ARPES experiment \cite{Bendounan2011}. It was attributed to a
surface resonance derived from the Shockley surface state of Pt(111), which is split on both sides of $\overline{\Gamma}$ by a large 
RB-type effect. This state, which for clean Pt(111) is located above E$_{\textmd{F}}$ in a hybridization gap, moves below E$_{\textmd{F}}$ as a result of the interaction with the
Ag overlayer. The shape, splitting, binding energy, and also the rather diffuse
intensity, all agree with the ARPES data for 3 ML from Ref. \onlinecite{Bendounan2011}. The Ag(111) $(1\times1)$ LEED pattern is first seen above
$3$ ML, and the RB-split band completely disappears above  $\Theta_{\textmd{Ag}}=4$\ ML. For larger 
Ag coverages the Ag(111) Shockley surface state is observed around $\overline{\Gamma}$ [Fig. 2(c)].

\begin{figure}[!b]
  \centering
  \includegraphics[width = 8.6 cm]{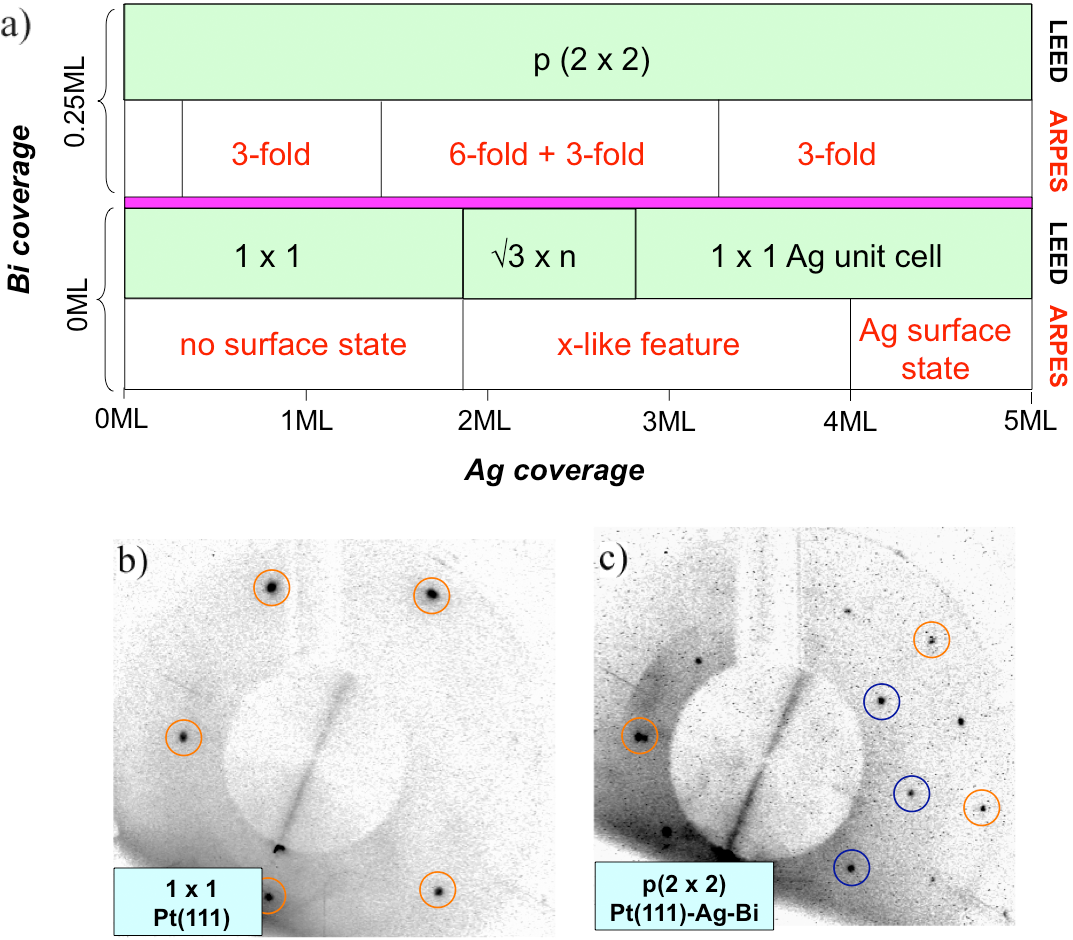}
  \caption{(color online) (a) A schematic phase diagram of the main ARPES and LEED
  results as a function of the Ag and Bi coverage. ARPES results
  for $\Theta_{\textmd{Bi}}=0$\ ML refer to the shallow Ag-induced state around
  $\overline{\Gamma}$. ARPES results
  for $\Theta_{\textmd{Bi}}=0.25$\ ML refer to the 
  interface states around $\overline{\Gamma}$. (b) LEED pattern at $90$ eV for a clean Pt(111) substrate.
  Orange (gray) circles denote $(1\times1)$ spots. (c) LEED pattern at $92$ eV after the
  deposition of Ag and Bi revealing a $p(2\times2)$ reconstruction. Orange (gray) circles denote $(1\times1)$ spots.
  Blue (black) circles denote $p(2\times2)$ spots.} \label{fig3}
\end{figure}

The deposition of bismuth  induces significant
changes in the LEED and ARPES signatures, summarized in Fig. 3. The Bi evaporation
source was calibrated using the characteristic LEED pattern of one ML of Bi on a pristine Ag (111) substrate  \cite{Toney1991}. A Bi coverage of
$\sim0.25$\ ML yields a sharp $p(2\times2)$ LEED pattern, irrespective of the thickness of the Ag layer, i.e both for the simple
$(1\times1)$ ($x<2$\ ML; $x>3$\ ML) and for the reconstructed $(\sqrt{3}\times n)$
($2<x<3$\ ML) Pt(111)-Ag interface. The $(2\times2)$ structure is never observed for the Bi-free
Pt(111)$-x$MLAg interface. 
As already mentioned, we obtained identical LEED and ARPES results even when
the deposition order of Ag and Bi was reversed, i.e. when Bi was directly evaporated on the Pt(111) substrate.
Further Bi evaporation up to $0.5$\ ML does not yield any new superstructure, but only results in a progressive deterioration of the $(2\times2)$ pattern. 
The data presented in the following refer to $\Theta_{\textmd{Bi}}=0.25$\ ML.

These obervations indicate that the $(2\times2)$ superstructure corresponds to a Bi-induced reconstruction  
where Bi most likely floats on top of the Ag layer. Bismuth therefore behaves as a surfactant in the layer-by-layer growth of Ag on Pt(111).
This is not surprising if one considers the well-known surfactant properties of Sb \cite{Oppo1993,Vegt1998,Scheuch1994} -- which is 
isoelectronic and has a smaller atomic radius --  and of Pb \cite{Camarero1994,Camarero1998,Passeggi2000}, which 
immediately precedes Bi in the periodic table.
On the other hand, we could never obtain the  $(\sqrt{3}\times\sqrt{3})$R30$^{\circ}$ pattern typical of the BiAg$_2$ surface 
alloy formed for 1/3 ML Bi coverage on the Ag(111) single crystal surface. Clearly, the strain-induced structural modifications at the Pt(111)-Ag
interface are large enough to modify the chemistry of the topmost Ag layer with respect to the pristine Ag(111) surface.
A detailed structural investigation, e.g. by surface x-ray diffraction, and first-principles total energy calculations could clarify this point.

 \begin{figure*}
  \centering
  \includegraphics[width = 11.4 cm]{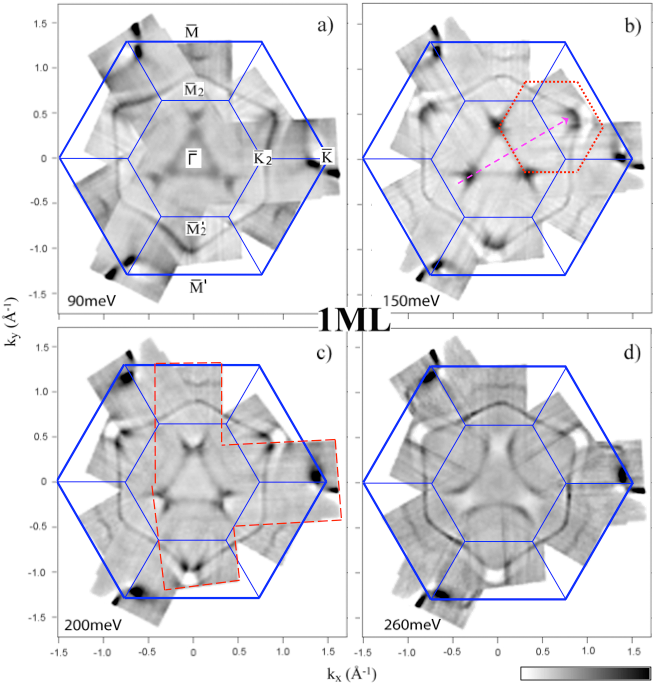}
  \caption{(color online) (a) -- (d) Constant energy ARPES intensity maps at four different binding energies for (Pt(111)$-1$ ML Ag$-$Bi$)-(2\times2)$.  
Data were collected in the region enclosed by the dashed lines in panel (c) and then symmetrized using a threefold axis. Bold and 
thin hexagons mark the $(1\times1)$ and $(2\times2)$ surface
Brillouin zones. The dotted hexagon follows the contour of a NFE paraboloid centered
at $\overline{\textmd M}_{2}'$ (see Appendix), which is only partially visible due to ARPES matrix elements. 
The dashed arrow in is a reciprocal lattice vector of the  $(2\times2)$ structure, connecting replicas of interface band features.
The second derivative of the photoemission intensity has been used to enhance the experimental features.
Intensity follows the attached gray-scale bar where signal-to-noise ratio increases from the left to the right.}
\label{fig4}
\end{figure*}
\begin{figure*}
  \centering
  \includegraphics[width = 9.8 cm]{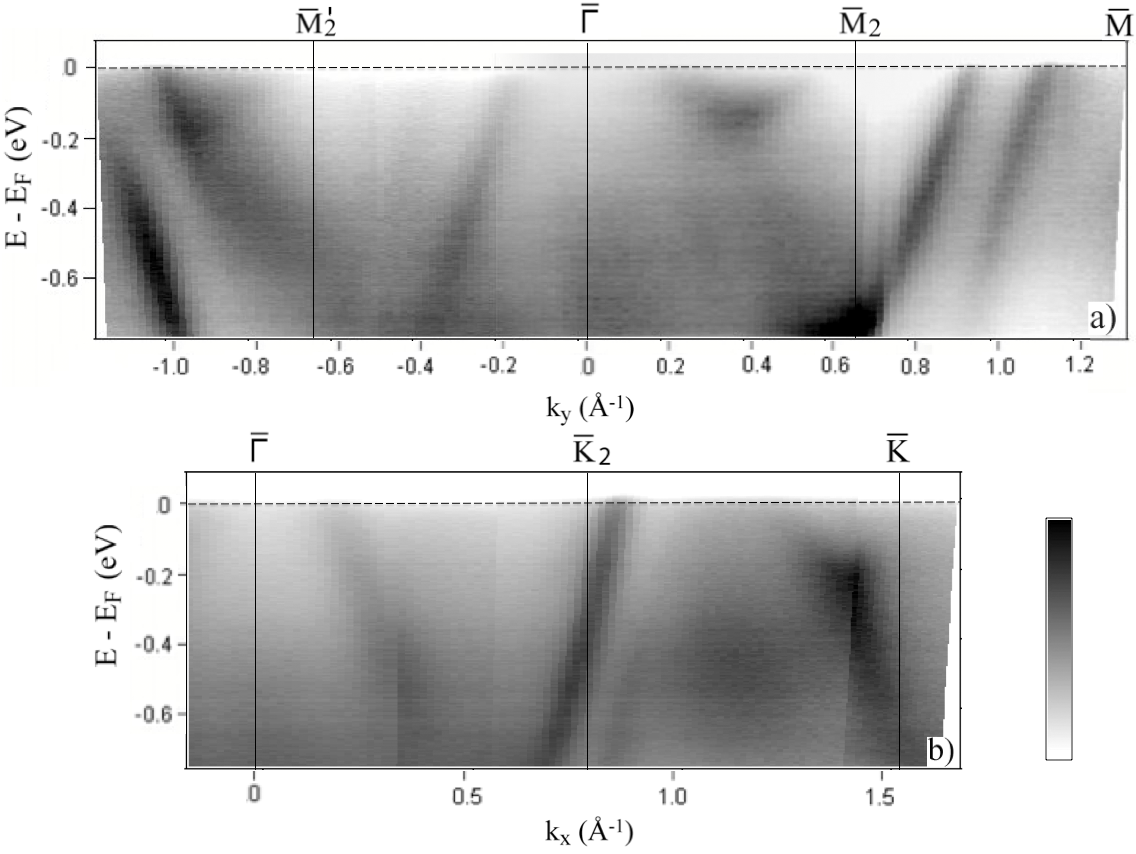}
  \caption{ARPES band dispersion for (Pt(111)$-1$ ML Ag$-$Bi$)-(2\times2)$ along the high-symmetry directions of the surface BZ.
  Intensity follows the attached gray-scale bar where signal-to-noise ratio increases from the bottom to the top.} \label{fig5}
\end{figure*}
\begin{figure*}
  \centering
  \includegraphics[width = 11.5 cm]{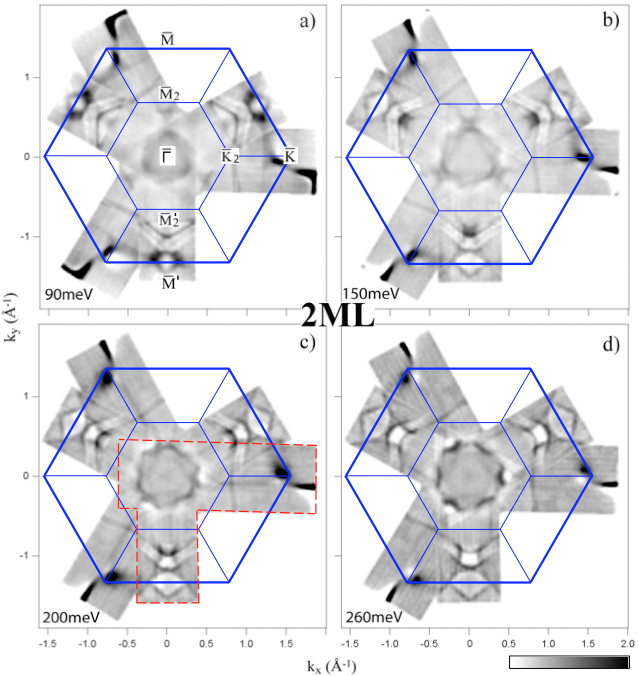}
  \caption{(color online) (a) -- (d) Constant energy ARPES intensity maps at four different binding energies for (Pt(111)$-2$ ML Ag$-$Bi$)-(2\times2)$. 
 Data were collected in the region enclosed by the dashed lines in panel (c) and then symmetrized using a threefold axis. 
 Bold and thin hexagons mark the $(1\times1)$ and $(2\times2)$ surface
Brillouin zones. The dashed arrow in (b) is a reciprocal lattice vector of the  $(2\times2)$ structure, connecting replicas of interface band features.
The second derivative of the photoemission intensity has been used to enhance the experimental features.
Intensity follows the attached gray-scale bar where signal-to-noise ratio increases from the left to the right.} \label{fig6}
\end{figure*}
\begin{figure*}
  \centering
  \includegraphics[width = 10.0 cm]{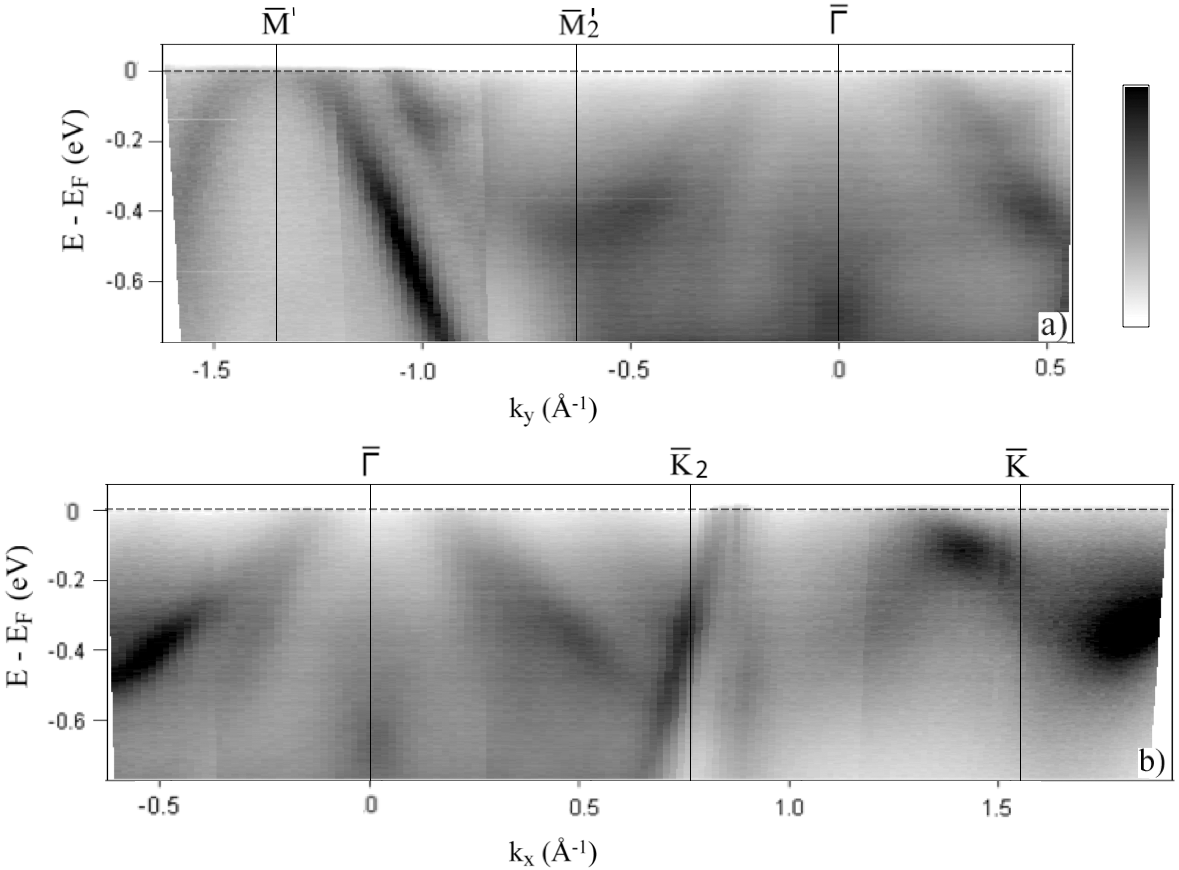}
  \caption{ARPES band dispersion for (Pt(111)$-2$ ML Ag$-$Bi$)-(2\times2)$ along the high-symmetry directions of the surface BZ.
  Intensity follows the attached gray-scale bar where signal-to-noise ratio increases from the bottom to the top.} \label{fig7}
\end{figure*}
\begin{figure*}
  \centering
  \includegraphics[width = 15.4 cm]{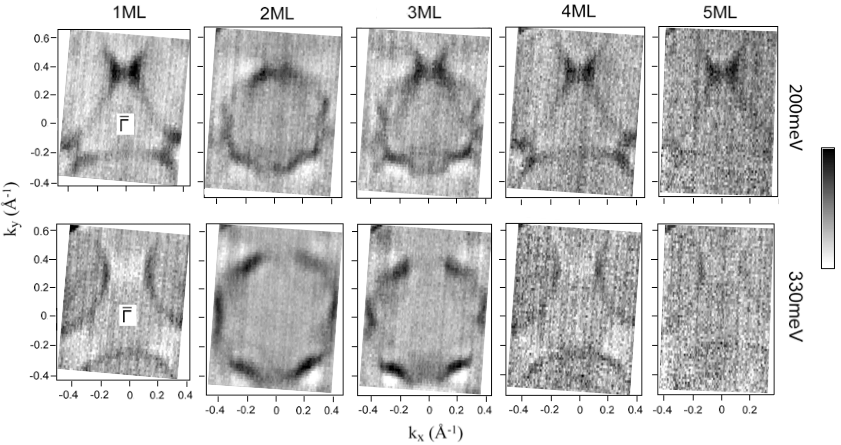}
  \caption{Pt(111)-Ag-Bi: Constant energy ARPES intensity maps at two binding energies for various Ag coverages
   limited to $k$-values around $\overline{\Gamma}$. They show the evolution from a threefold
   rotational symmetry at $1$ ML, to a superposition of threefold and sixfold at $2$ and $3$ ML, and back to threefold at $4$ ML.
  The maps were not symmetrized. The second derivative of the photoemission intensity has been used to enhance the experimental features.
Intensity follows the attached gray-scale bar where signal-to-noise ratio increases from the bottom to the top.}
   \label{fig8}
\end{figure*}

ARPES data for the $\Theta_{\textmd{Ag}}=1$\ ML case are illustrated in Figures 4 and 5. Figure 4 (a)$-$(d) are CE intensity maps,
covering a range of $k$ space similar to that of Fig. 1, for four binding energies between 90 meV and 260 meV. 
Figure 5 (a) and (b) shows two corresponding energy vs. wave vector intensity plots along the 
$\overline{\textmd{M}}'\overline{\Gamma\textmd{M}}$ ($k_x=0$) and $\overline{\Gamma\textmd{K}}$ ($k_y=0$) high symmetry directions
in the surface BZ. 
From a comparison with Fig. 1, one can identify in both figures the Pt-derived states, namely the bands crossing E$_{\textmd F}$
at $0.9$ \AA$^{-1}$ and at $1.1$ \AA$^{-1}$ in the  $\overline{\Gamma\textmd{M}}$ direction. Remarkably, the CE contours of
the former exhibit a distorted hexagonal shape, with threefold symmetry, rather than the sixfold symmetric shape
of clean Pt(111) reported in Ref. \onlinecite{Di1991} and inferred by the results of Fig. 1. In addition of the slightly different photon energy
of the two studies ($24$ eV vs. $21.2$ eV), we propose other factors which may influence the overall shape of the hexagonal contour.
We speculate that with the formation of the Ag+Bi overlayer this state has acquired a stronger bulk character. Therefore, it may reflect the
threefold symmetry of the bulk in a stronger way. Alternatively, the change in work function -- and consequently in the inner potential V$_0$ -- could be large enough
to spoil the accidental condition at the origin of the apparent sixfold symmetry reported for clean Pt(111). A different possible origin of the threefold modulation 
might be the interaction of the Pt-derived surface resonance with electronic states having a threefold symmetry. Such states will be described in the following. 

Indeed a further analysis of Fig. 4 and 5 reveals electronic states which have no counterpart in the electronic structure of the Pt(111) substrate or the Pt(111)-Ag interface. 
New CE contours appear centered around $\overline{\Gamma}$. They evolve from a nearly triangular shape [Fig. 4(a),(b)] to three 
disconnected arcs [Fig. 4(c),(d)] with increasing binding energy. Their origin must be found in three identical bands  upward dispersing from
the three equivalent $\overline{\textmd M}_{2}'$ points of the $(2\times2)$ surface BZ, and crossing E$_{\textmd F}$ near 
$\overline{\Gamma}$. This speculation is further developed by a phenomenological model in the Appendix of the present work. One of these bands 
is readily visible along $\overline{\Gamma\textmd M}_{2}'$ and $\overline{\Gamma\textmd K}$ respectively in Fig. 5 (a) and (b). The nearly circular 
CE contours [Fig. 4(d)], which are only partially visible due to ARPES matrix elements, indicate that these bands have a nearly free-electron-like 
character near their origin. At larger wave vectors they experience a stronger effect of the lattice potential, and the 
CE contours become hexagonal. One of them is outlined by the dotted hexagon in Fig. 4(b). The three bands 
cross at a binding energy of $150$ meV, yielding a triangular contour with strong intensity at the vertices at this energy. 
The shallow pocket along $\overline{\Gamma\textmd M}_{2}$ is a signature of the intersecting states. At the same time, band splitting is evidenced
above the vertices of the triangular contour in Fig. 4(a). The splitting demonstrates that the interaction between the corresponding states is non negligible for other
wave vectors. In the Appendix, we present a phenomenological model which quantifies the hybridization of these bands.
The new interface states feel the $(2\times2)$ periodicity of the system. This is appreciated most clearly in Fig. 4(b). Replicas of the 
three intense crossings around $\overline{\Gamma}$, connected by reciprocal lattice vectors of the superstructure, are seen in the adjacent surface BZs, 
overlapping the Pt-derived states. In summary, all electronic states of the (Pt(111)$-$1\ ML Ag$-$Bi$)-(2\times2)$ system exhibit the threefold rotational 
symmetry of the (111) surface of an fcc~lattice.

ARPES data for the $\Theta_{\textmd{Ag}}=2$\ ML case are illustrated by the CE intensity maps of Fig. 6 and by the corresponding energy vs. wave vector intensity plots of Fig. 7. The Pt-derived bands and the interface states discussed above can still be identified, but a new state appears at this coverage. Its CE contour exhibits a sixfold symmetry, most clearly visible in Fig. 6 (d). It overlaps with and partially masks the triangular contour of the 1\ ML case. The dispersion of this new state can be identified in the intensity maps of Fig. 7. In the $\overline{\Gamma\textmd M}_{2}'$ direction [Fig. 7(a)] it has a minimum at the $\overline{\textmd M}_{2}'$ point at a binding energy of $\sim$0.5 eV. Along
$\overline{\Gamma\textmd K}$ [Fig. 7 (b)] its Fermi level crossing is essentially degenerate with that of the 1\ ML state, but its Fermi velocity is smaller. 

Details of the evolution of the Bi-induced bands as a function of $\Theta_{\textmd{Ag}}$ are illustrated by Fig. 8.
The figure presents second-derivative maps of the ARPES intensity around the $\overline{\Gamma}$ point
for 1\ ML\ $\leq\Theta_{\textmd{Ag}}\leq$\ 5\ ML, and at two binding energies ($200$ meV and $330$ meV).
The data were not symmetrized in order to prevent possible artifacts from the symmetrization procedure.
At  $\Theta_{\textmd{Ag}}=1$ ML (left vertical panel), the CE contours exhibit a clear threefold symmetry.
The possibility that ARPES matrix element effects determine the shape of the triangular contour has been
ruled out by varying the experimental geometry. The second vertical panel shows that at $2$ ML dominant (especially at $330$ meV)
sixfold contours are superimposed on weaker traces of the triangular contours, as already noticed in Fig. 6.
The two contours coexist also at $3$ ML, with a more balanced intensity. At $4$ ML  
the sixfold contour is missing and the only Bi-induced interface state around $\overline{\Gamma}$
has a threefold symmetry. At $5$ ML the threefold contour is still visible, but more blurred. This reflects
disorder associated with the formation of three-dimensional Ag islands \cite{Roder1997}. 
The sequence of images of Fig. 8 clearly illustrates the
threefold\ $\rightarrow$\ threefold$+$sixfold\ $\rightarrow$\ threefold evolution of the CE contours, i.e. a re-entrant behavior
of the rotational symmetry of the Bi-induced interface states as a function of the Ag coverage.

\section{IV. Discussion}

The properties of the electronic states in a solid are strongly constrained by symmetry requirements.
In the absence of a magnetic field, time-reversal (TR) symmetry requires that:
\begin{equation}
E_{k,\uparrow (\downarrow)}=E_{-k,\downarrow (\uparrow)}\ ,
\label{E1}
\end{equation}
where the arrows stand for the spin-polarization. In the limit of a
vanishing energy separation between the two spin states, Eq.
(\ref{E1}) reduces to the simpler $E_{k}=E_{-k}$. In a 2D close-packed system
with a sixfold unit cell, irrespective of the magnitude of the
spin-separation, a spin-integrated technique such as ARPES
yield electronic contours of hexagonal in-plane symmetry
satisfying (\ref{E1}). Even if the 2D system only admits a threefold rotation axis, Eq. (\ref{E1})
still requires that the CE contours of the electronic structure exhibit a sixfold symmetry (\ref{E1}) \cite{Premper2007,Frantzeskakis2010}.
Therefore, as already pointed out, surface states cannot exhibit a threefold rotational symmetry, because
this would be incompatible with TR symmetry.

Unlike surface states, bulk states are characterized by a well-defined perpendicular wave vector $k_{\perp}$, and
(\ref{E1}) applies to the 3D $\vec{k}$-vector. The combined effect of (\ref{E1}) and of
a threefold rotation axis yields a 3D band dispersion with an overall
threefold symmetry. Prime examples are the bulk electronic
structures of fcc metals \cite{Ashcroft}. ARPES is only sensitive to the in-plane component 
of $\vec{k}$, and maps the full 3D dispersion onto the surface BZ.
Therefore, in an ARPES measurement from the (111) surface of an fcc system, the threefold symmetry of the 
bulk states coexists with the hexagonal surface BZ, e.g. for the well-studied case of Cu(111)
\cite{Aebi1994,Baumberger2003}.
For values of the surface wave vector outside the projected bulk gaps, there are no true
surface states but only surface resonances, which hybridize significantly with the
continuum of bulk states. In an fcc system this typically yields a threefold modulation in their momentum
distributions. This is e.g. the case of the hole pockets at the Sb(111) surface \cite{Sugawara2006}. 
Our observation of sixfold CE contours around $2$ ML suggests that the crystal structure has also acquired
a sixfold symmetry at this coverage. 
This is indeed consistent with the scenario  of the Pt(111)-Ag interface developed
from structural investigations.

As already mentioned in the introduction, the first monolayer of Ag on Pt(111) grows heteroepitaxially
conserving the fcc stacking of the substrate. The $4.3$\% difference in the
lattice constants of the two materials yields a
coherently strained commensurate
overlayer. Strain is relieved with the completion of the second monolayer. At the annealing temperature ($400$ K) used in the present study, the SI phase is formed \cite{Brune1994}.
Moreover, at this temperature there is no intermixing between Pt and Ag atoms \cite{RoderPRL1993}. In the $2$\ ML SI phase, regions with fcc and hcp stacking coexist. The majority domains have been
assigned to hcp stacking both by STM \cite{Roder1997} and photoelectron diffraction \cite{Rangelov1995}, although an earlier STM study had reported that fcc stacking is dominant \cite{Brune1994}. Both experiments
agree that with further Ag deposition the fcc
stacking of the substrate is resumed, and that the growth is mainly two-dimensional at RT
up to a critical thickness of $6-9$ ML
\cite{Roder1997,Rangelov1995}.

The fcc stacking implies a threefold symmetry. By contrast, the symmetry resulting from the hcp stacking is
sixfold. The wavefunctions of the interface resonances certainly extend into the bulk by at least $3$ ML, which is enough
to ``feel'' the difference between the two different stacking sequences. Hcp
domains exist on the uppermost layers only after the deposition of
$2$\ ML of Ag, and this is consistent with our observation of coexisting threefold and sixfold CE contours at this
coverage. As more Ag is deposited, the fcc stacking of
the uppermost layers is reflected in the electronic structure by
the dominance of the CE contour with a triangular shape. Indeed, the orientation of the
triangular contour is consistent with the structural reflection symmetry of an 
fcc slab. Therefore, the sequence
threefold\ $\rightarrow$\ threefold$+$sixfold\ $\rightarrow$\ threefold 
finds a natural explanation in the growth mechanism of
Ag on Pt(111). In other words, ARPES successfully reveals the symmetry of the
growth domains and the Ag-coverage-dependent transition. Following the
above line of reasoning, our results suggests that hcp sites are the
majority domains at $\Theta_{\textmd{Ag}}\ = 2$ ML because they evidence a predominant sixfold symmetry for $0.25$ ML
Bi on $2$ ML Ag. This conclusion is in good agreement with the structural analysis presented in Refs. \onlinecite{Roder1997} and \onlinecite{Rangelov1995}.

Apart from the experimental geometry, ARPES matrix element effects can be introduced by the scattering process of the photo-emitted electrons.
These final state effects are the analog of a LEED $IV$ or an X-Ray Photoelectron Diffraction (XPD) experiment.
In the latter case, the associated modulation of intensity for a core-level photo-electron
can be of the order of several tens of \%. This would introduce a threefold intensity modulation
for states with different symmetries (e.g. circular, hexagonal). In the present work, threefold and sixfold contours are rather different and coexist on the same sample
for $\Theta_{\textmd{Ag}}\ = 2-3$ ML (Fig. 8). There is no sign of interaction of the corresponding states but rather a superposition of their contours.
Therefore, they must originate from different domains of the sample, which are naturally attributed to the fcc and hcp domains.
In line with these arguments, a sixfold contour was never evidenced on interfaces with an fcc(111) structure (i.e. $\Theta_{\textmd{Ag}}\ = 1$ ML
and $\Theta_{\textmd{Ag}}\ \geqslant 4$ ML).

Concerning the Bi atoms, our data suggests that they reside on the topmost layer. They
induce the formation of a long-range reconstruction by
preferential ordering at $(2\times2)$ sites, and do
not affect the structural symmetry of the underlying
Pt(111)-Ag interface. Nonetheless, further structural meaurements are necessary to fully characterize the Pt(111)-Ag-Bi trilayer.

Is there any signature of RB-type splitting on the studied interfaces? As discussed in Section III, our results on the Bi-free interface are in
good agreement with Bendounan \textit{et al. }\cite{Bendounan2011}. Fig. 2 evidences the unoccupied surface state of Pt(111) \cite{Wiebe2005}
which shifts below $E_{F}$ after Ag deposition of $2-3$ ML and presents an enhanced RB splitting. The latter was attributed to
multiple scattering between bulk and surface \cite{Bendounan2011}. Our results suggest that there is still no alloying after the deposition of Bi.
Therefore, it is multiple scattering between bulk and surface, rather than a strong in-plane potential gradient who could act as a potential
mechanism for a strong RB-splitting of surface states. On the other hand, the Pt(111)-Ag surface resonance disappears after Bi deposition and is 
replaced by new electronic states with a finite bulk extension. The latter do not present any direct evidence of RB splitting. The experimental data for  
Pt(111)$-1$ ML Ag$-$Bi is supported by a phenomenological NFE model (Appendix) where the main band structure features have been reproduced
by a spin-free simulation. The situation is less clear for Pt(111)$-2$ ML Ag$-$Bi where new states appear. We hope that our results will motivate
relativistic \textit{ab initio} studies which can readily identify the spin character of the Bi-induced states.  

\section{V. Conclusions}

We presented a detailed ARPES investigation of an ordered
Pt(111)-Ag-Bi trilayer system.
In the studied Ag coverage range $1\ \leq\Theta_{\textmd{Ag}}$\ $\leq$\ $5$ ML we did not observe the expected  formation 
of the BiAg$_2$ ``Rashba" surface alloy with  periodicity $(\sqrt{3}\times\sqrt{3})$R30$^{\circ}$, characteristic of the Bi-Ag interface on a pristine Ag(111) surface.
We observed instead a novel $p(2\times2)$ phase, where the Bi atoms most likely 
float on top of the Ag interlayer yielding strong interface states which change with the Ag coverage.
The rotational symmetry of their CE contours evolves from threefold,
to a superposition of threefold and sixfold, and finally back to threefold.
This evolution is consistent with the accepted model for the growth of Ag on the
Pt(111) substrate, namely with a strain-induced transition at $2$ ML Ag coverage.
These results illustrate the consequences of the structural changes on the character
of the electronic structure. They also show that an analysis of the symmetry properties
of the electronic states of a system may provide valuable insight into its structural properties.

\section{Acknowledgements}

We thank J. Audet for experimental contributions during the early
stage of this work. E.F. acknowledges the financial support of the
Alexander S. Onassis Public Benefit Foundation. This research was
supported by the Swiss NSF and the NCCR MaNEP.

\section{Appendix}

We have used a simple model to fit the experimental band structure of the Pt(111)$-1$ ML Ag$-$Bi trilayer around
$\overline{\Gamma}$. Our aim is to quantify the interaction of the electronic states with threefold symmetry. 
Following the results presented in Section III, we have used NFE paraboloids
centered at the three equivalent $\overline{\textmd M}_{2}'$ points of the $1^{\textmd{st}}$ $p(2\times2)$
surface BZ. The hybridization strength $V$ is included as an $ad$ $hoc$ parameter in the off-diagonal elements
of the resulting $3\times3$ Hamiltonian matrix. The model dispersion can be inferred by the matrix eigenvalues and is plotted in Fig. 9.
\begin{figure}
  \centering
  \includegraphics[width = 8.4 cm]{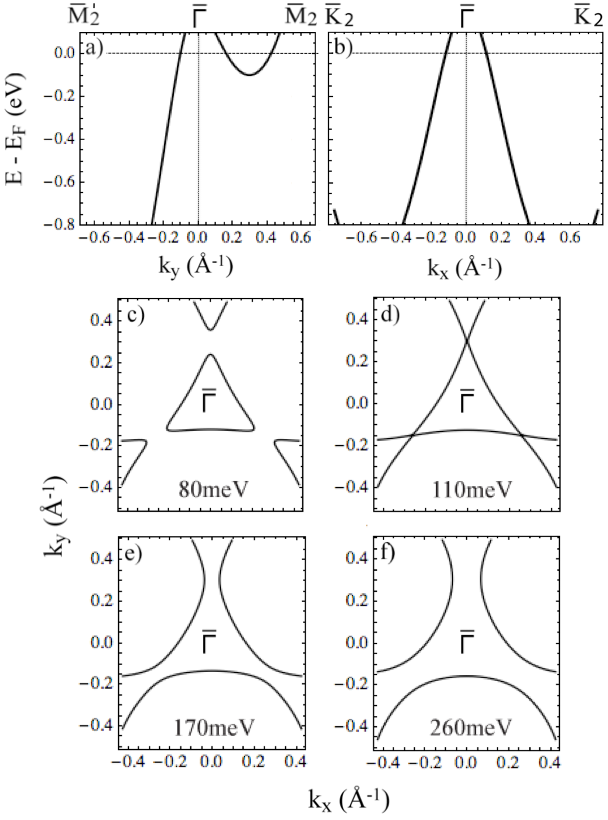}
  \caption{(a),(b) Model band dispersion for (Pt(111)$-$1\ ML Ag$-$Bi), along the high-symmetry directions of the $(2\times2)$ surface BZ.
  (c) -- (f) Constant energy contours at four different binding energies. The results have been acquired for NFE states with $m^{*}=0.53m_{e}$
  and an interaction parameter $|V|=0.6$ eV.} \label{fig9}
\end{figure}

Results are presented for a NFE effective mass of $0.53m_{e}$ and a hybridization parameter $|V|$ of $0.6$ eV.
The model dispersion along the two high-symmetry directions reproduces the main experimental features observed in Fig. 5.
Moreover, CE contours around $\overline{\Gamma}$ show a remarkable one-to-one correspondence with the experimental
maps of Fig. 4. Despite its simplicity, the model can capture the triangular contour around $\overline{\Gamma}$, the pockets around
the $\overline{\textmd M}_{2}$ points and the deviation of the momentum distributions from perfect circles with decreasing binding energy.

This is a purely phenomenological description and cannot be extended to higher $k$ values and different experimental states.
Moreover, as a tradeoff to its simplicity, there are a few quantitative discrepancies from the experimental values. Nevertheless,
it presents a solid support to the claim that a finite hybridization is necessary for the observed experimental dispersion.

\end{document}